\documentclass[useAMS,usenatbib]{mn2e}
\usepackage{graphicx}
\usepackage[totalwidth=480pt, totalheight=680pt]{geometry}
\pdfoutput=1

\title[The Eagle Galaxy at z=0.77]{Oxford SWIFT IFS and multi-wavelength observations of the Eagle galaxy at z=0.77}

\author[S. A. Kassin et al.]{
Susan A. Kassin,$^{1,2}$\thanks{E-mail: susan.kassin@nasa.gov} 
L. Fogarty,$^1$
T. Goodsall,$^{1,3}$
F. J. Clarke,$^1$
R. W. C. Houghton,$^1$
\newauthor 
G. Salter,$^1$
N. Thatte,$^1$
M. Tecza,$^1$
Roger L. Davies,$^1$
Benjamin J. Weiner,$^4$
\newauthor 
C. N. A. Willmer,$^4$
Samir Salim,$^5$
Michael C. Cooper,$^6$
Jeffrey A. Newman,$^7$
\newauthor 
Kevin Bundy,$^8$
C. J. Conselice,$^9$
A. M. Koekemoer,$^{10}$
Lihwai Lin,$^{11}$
\newauthor 
Leonidas A. Moustakas,$^3$
Tao Wang$^{12}$\\
$^1$ Sub-Department of Astrophysics, University of Oxford, Denys Wilkinson Building, Keble Road, Oxford OX1 3RH, UK\\
$^2$ currently: Astrophysics Science Division, Goddard Space Flight Center, Code 665, Greenbelt, MD 20771, USA\\
$^3$ Jet Propulsion Laboratory, California Institute of Technology, 4800 Oak Grove Drive, MS 169-327, Pasadena, CA 91109, USA\\
$^4$ Steward Observatory, 933 N. Cherry St., University of Arizona, Tucson, AZ 85721, USA\\
$^5$ Department of Astronomy, Indiana University, Bloomington, IN 47404, USA\\
$^6$ Center for Galaxy Evolution, Department of Physics and Astronomy, University of California, Irvine,\\ 4129 Frederick Reines Hall, Irvine, CA 92697, USA; Hubble Fellow\\
$^7$ Department of Physics and Astronomy, University of Pittsburgh, 3941 O'Hara Street, Pittsburgh, PA 1526, USA\\
$^8$ Astronomy Department, University of California, Berkeley, CA 94705, USA; Hubble Fellow\\
$^9$ University of Nottingham, School of Physics and Astronomy, Nottingham NG7 2RD, UK\\
$^{10}$ Space Telescope Science Institute, 3700 San Martin Drive, Baltimore, MD 21218, USA\\
$^{11}$ Institute of Astronomy \& Astrophysics, Academia Sinica, Taipei 106, Taiwan\\
$^{12}$ Harvard-Smithsonian Center for Astrophysics, 60 Garden Street, Cambridge, MA 02138, USA\\
}

\begin{document}

\maketitle

\begin{abstract}
The `Eagle' galaxy at a redshift of 0.77 is studied with the Oxford Short Wavelength Integral Field 
Spectrograph (SWIFT) and multi-wavelength data from the All-wavelength Extended Groth 
strip International Survey (AEGIS).  It was chosen from AEGIS because of the bright and extended
emission in its slit spectrum.  Three dimensional kinematic maps of the Eagle reveal a gradient
in velocity dispersion which spans $35-75\pm10$ km s$^{-1}$ and a rotation velocity of 
$25 \pm 5$ km s$^{-1}$ uncorrected for inclination.  {\it Hubble Space Telescope} images
suggest it is close to face-on.  In comparison with galaxies from AEGIS at similar 
redshifts, the Eagle is extremely bright and blue in the rest-frame optical, highly star-forming, 
dominated by unobscured star-formation, and has a low metallicity for its size.  
This is consistent with its selection.  The Eagle is likely undergoing a major merger and is
caught in the early stage of a star-burst when it has not yet experienced metal enrichment
or formed the mass of dust typically found in star-forming galaxies.
\end{abstract}


\section{Introduction}
Large redshift surveys at $0 <z<1.2$ have revealed that the population of galaxies is
divided into red and blue rest-frame colours, and have measured the evolution of
these populations \citep[e.g.,][]{stra,bell,fab7}.
In particular, blue galaxies in the past were brighter 
by $B \simeq 1.3$ mag \citep[e.g.,][]{bell, cnaw}, more highly star-forming
\citep[e.g.,][]{kai}, and more morphologically irregular \citep[e.g.,][]{abra, oesc}.
In addition, blue galaxies have an increasing contribution to their kinematics from 
disordered motions the further one looks back in time \citep{wei1, kas7}.  

These disordered motions, quantified by an integrated velocity dispersion, appear to 
play an important role in galaxy kinematics \citep[e.g.,][]{wei2, kas7, cres, covi, fors9, puec10, lemo1, lemo2}.
A study of $\sim 550$ galaxies with slit spectroscopy over $0.1<z<1.2$ has shown that, for galaxies with stellar
masses greater than $10^{10}$ M$_{\odot}$, there is increasing
scatter in the Tully-Fisher relation (which relates the rotation velocities of galaxies 
to their stellar masses or magnitudes) to $z=1.2$ \citep{kas7}.    This scatter is dominated by galaxies with
disturbed morphologies \citep{kas7}.  A large scatter in the
Tully-Fisher relation is also found for galaxies in integral field
spectrograph (IFS) studies at $z\sim 0.6$ \citep{flor,puec8,puec10a} 
and $z \sim 2-3$ \citep[e.g.,][]{law, cres, lemo2}.

When a kinematic estimator which incorporates
both rotation velocity ($V$) and integrated velocity dispersion ($\sigma$) is adopted, 
{\small $S_{K} \equiv \sqrt{K V^2 + \sigma^2}$} with K=0.5 \citep{wei1}, the resulting
relation with stellar mass has remarkably small scatter at all redshifts \citep{kas7, cres, covi, puec10, lemo1, lemo2}.  This is
such that disturbed galaxies have higher values of $\sigma$ than
normal disc-like galaxies.  Undisturbed disc galaxies in the local Universe have $\sigma$ values that range from
10--35 km s$^{-1}$ and an average value of $\sim$20--25 km s$^{-1}$ \citep{ghasp} which
are due to the relative motions of individual gas clouds in spiral arms or a thick disc.  The 
higher $\sigma$ values found for many high redshift galaxies likely represent effective velocity dispersions caused by the blurring of 
velocity gradients on scales at or below the seeing limit \citep{wei1, kas7}.  These may not even have
a preferred plane.

The nature of these kinematically peculiar objects cannot 
be fully determined from slit spectroscopy alone since it is unable to probe the full 3D kinematics of galaxies.
3D integral field observations can give further clues as to
whether the high $\sigma$ values are driven by phenomena such as rotation,
major or minor merger activity, and/or violent star-formation processes.

The advent of IFSs on large telescopes has allowed for detailed studies of 3D galaxy kinematics
over $0.4 \la z \la 3$ \citep[e.g.,][]{fors6, fors9, puec7, yang, vans, wrig, law, epin, lemo1, lemo2}.
These studies have all found galaxies with large $\sigma$ values
and attribute it to star-formation and major and minor merger activity, when
active galactic nuclei are not present.  However, these samples are still
small ($\la 100$) and are typically too bright to be representative of 
typical galaxies, although there are exceptions \citep[e.g.,][]{puec8, yang, law}.

A representative sample of 68 galaxies at redshifts $0.4 < z < 0.74$ has been studied with
the GIRAFFE IFS with the Intermediate MAss Galaxy Evolution
Sequence (IMAGES) Survey.
Two main results of this survey are  the discovery of a low fraction of rotating disc galaxies \citep{flor, neic,yang} and 
the finding that more distant galaxies have smaller ratios of $V$ to $\sigma$ than
galaxies at lower redshifts \citep{puec7}.  These results are consistent with the findings
of slit based studies of larger samples of $\sim$500--1000 galaxies \citep{wei1,wei2,kas7}.
Studies  of IMAGES galaxies generally attribute disturbed kinematics to major mergers.
 
It is highly desirable to study more galaxies with IFSs at $z\sim1$, an epoch probed by large redshift  
surveys but with few IFS observations.
In this paper, IFS observations of the 3D kinematics of a galaxy at $z=0.7686$ with a high $\sigma$ value
are presented along with a suite of multi-wavelength data from the
All-wavelength Extended Groth strip International Survey \citep[AEGIS,][]{aegis}.
These data allow for the galaxy to be understood in terms of the general population of
galaxies at its epoch and offer insight into a likely cause for its high $\sigma$ value.
A $\Lambda$CDM cosmology is adopted throughout: $H=70$ km s$^{-1}$ Mpc$^{-1}$, $\Omega_m=0.3$, $\Omega_\Lambda=0.7$.
All logarithms are base 10, and all magnitudes are on the AB system.

\section{The Eagle galaxy}

\subsection{Selection}

\begin{figure*}
\includegraphics[scale=1.7]{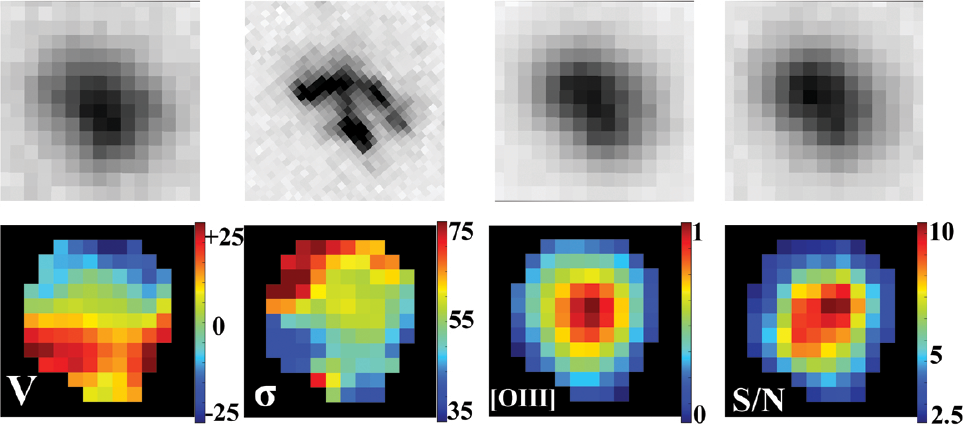}
\caption{Broad-band and SWIFT IFS observations of the Eagle galaxy are shown in the top
and bottom panels, respectively.  All panels are 3\arcsec\ on a side and are oriented 
such that north is up and east is to the left.  Top row: Broad-band images at $B$ 
(CFHT/CFH12K; 0.21\arcsec pixel$^{-1}$),  $V$ ($Hubble/ACS$; 0.1\arcsec pixel$^{-1}$), 
$R$ (CFHT), and $I$ (CFHT) are shown from left to right.  The image FWHM for the 
CFHT images was $\sim 0.7$\arcsec.
Bottom row: Maps of $V$, $\sigma$, [OIII] line emission, and signal-to-noise ratio for the [OIII] line
are shown from left to right.  These maps have been smoothed with a 
Gaussian with a $\sigma$ of 1 spaxel (0.235\arcsec),
which is much smaller than the $\sim 1$\arcsec\ image FWHM during the observations. 
Due to similar image FWHMs, the most direct comparison can be made between the SWIFT observations and 
the broad-band images from CFHT. 
\label{fig:SWIFT}}
\end{figure*}

The Eagle galaxy, DEEP2 ID 13019195 ($\alpha$ 14:19:39.3, $\delta$ +52:55:48.8, J2000), was selected from the AEGIS Survey \citep[for which the DEEP2 Survey
provides redshifts and spectra;][]{davi} to be observed with the Oxford Short Wavelength Integral Field 
Spectrograph \citep[SWIFT;][]{swift}.  The Eagle galaxy is so named for its distinctive appearance in 
{\it Hubble Space Telescope} images (Figures~\ref{fig:SWIFT} and \ref{fig:DEIMOS}).  
It was primarily selected from the AEGIS Survey to have very bright and extended line emission 
($\ga 10^{-16}$ ergs s$^{-1}$ cm$^{-2}$ and $\ga 2$\arcsec\, from a visual inspection of the spectrum, respectively) 
which falls into the SWIFT bandpass.  The aim of these requirements was to provide a target which would result
in observations with a high signal-to-noise ratio and at least two independent spatial resolution elements.
The [OIII] $\lambda 5007$ emission line in the DEEP2 spectrum of the 
Eagle has an integrated flux of $9.6 (\pm 2.5) \times 10^{-16}$ ergs s$^{-1}$ cm$^{-2}$ 
and a spatial extent of $\ga 2$\arcsec from a visual inspection.  

The target was further required to have at least one emission line not contaminated by sky lines, 
a redshift greater than $\simeq 0.7$,
an inclination $\la 70$\degr, and a significant $\sigma$ ($>40$ km s$^{-1}$).  
The limit on inclination allows for
kinematics to be studied across the face of the galaxy without significant projection effects,
and the requirement on $\sigma$ allows for an investigation into the cause of the
unusually high $\sigma$ values found for galaxies at high redshift. 

\subsection{Hubble morphology \& size} 
\subsubsection{Morphology}

$Hubble$/ACS $V$-band and $V + I$-band images are shown in 
Figures~\ref{fig:SWIFT} and \ref{fig:DEIMOS}, respectively.  Quantitative morphological analyses were
performed on the $I$-band image with the CAS and Gini/M$_{20}$ systems
(\citealt{cas} and \citealt{gini}, respectively).  Both classify the Eagle as a major merger galaxy.
Specifically, the Eagle has an asymmetry (A) value of 0.32 \citep{lin}, and although major mergers
are defined in CAS as galaxies with A$\ge$0.35, once the signal-to-noise in the ACS image
is taken into account, it is classified as such.  In addition, \citet{lin} classify 
galaxies with A$>$0.25 as major mergers.  The Eagle 
has Gini and M$_{20}$ values of 0.46 and -0.91, respectively \citep{lotz}.  These qualify 
it as a major merger according to the definition in
\citet{lotz}: Gini $> -0.14$M$_{20}$ + 0.33.  

Furthermore, according to the visual classification of $Hubble$  images in \citet{kas7}
which categorised galaxies as `normal,' `disturbed', or `compact', the Eagle is identified as `disturbed.'
This is mainly due to outer ispohotes which are not elliptical and an asymmetric three-armed structure, both of
which are inconsistent with the morphology of an undisturbed disc galaxy.
In summary, quantitative and qualitative morphological classifications of the
Eagle are consistent with it being a major merger galaxy and inconsistent with it being a disc galaxy.

\subsubsection{Size}

It is difficult to measure the sizes of high redshift galaxies which are 
not always smooth or elliptical.  In order to avoid biases due to disturbed
morphologies, we look to measure a constant fraction of the total galaxy light, 
independent of the morphology or orientation of the object.  The Petrosian radius 
comes closest to this ideal \citep[e.g.,][]{petro, bers, mass}.  
Petrosian radii measured in elliptical apertures are adopted from \citet{lotz}.  The
radii were measured from $I$-band
$Hubble$ images using the SExtractor software \citep{sext} for $\eta = 0.2$.  
The Petrosian radius of the Eagle is $1.2\arcsec \pm 0.2\arcsec$, which corresponds to 8.9 kpc at its redshift,
and the position angle of the major axis is 49\degr (east of north).

\section{SWIFT observations and data reduction}

The Eagle Galaxy was observed under clear conditions with SWIFT mounted on the Hale Telescope
at Palomar Observatory on May $8$th, $9$th, and $10$th of $2009$.  Observations were performed in natural seeing mode in the coarsest
spaxel scale of 0.235\arcsec, which results in a field of view of 10.3\arcsec $\times$ 20.9\arcsec.
The spectral resolution at the observed wavelength of 8855\AA\ is $R\sim4000$, which results in an instrumental broadening of 
$\simeq 1.2$ \AA\, or 41 km s$^{-1}$ (Gaussian $\sigma = 21$ km s$^{-1}$), as measured from fits of line 
profiles to isolated sky lines near the observed wavelength.
To obtain an accurate pointing, the telescope was offset blindly from a nearby bright star 
to the galaxy.  The large field of view allowed for dithering on source, and 
integrations of 900s each were taken.  During this commissioning run we were unable to guide the telescope, 
and relied on tracking alone.  A total of 6 exposures which had the best delivered image quality, or image FWHM,
were chosen to create the final cube, resulting in a total exposure time of 1.5 hours.  The image FWHM
was $\sim 1$\arcsec\, for these observations; they were taken at the beginnings of the
nights of May $9$th and $10$th.

The SWIFT data were reduced with custom-purpose software written by R. Houghton and T. Goodsall, based on the
{\tt spred} pipeline for the SINFONI IFS \citep{spred1, spred}.  The SWIFT reduction procedure is described in detail in \citet{arp}, and
is briefly reviewed here. 
The galaxy was observed in one of the two CCDs, so only data from that CCD was processed.  All
frames were reduced in the same manner.  First the bias was removed, the overscan region trimmed, and
cosmic rays removed with the L.A.Cosmic algorithm \citep{lacos}.  A wavelength solution was 
found from Ar and Ne arc lamp exposures taken at the beginning of the observing run.
Any flexure in wavelength between observations was quantified by measuring the shift
of sky lines close to the [OIII] line in wavelength.  For 2 of the 6 exposures used, there was flexure 
in the wavelength axis of 0.46 and 0.49 of a pixel (1 pixel = 1\AA).  These cubes were shifted 
in wavelength to align them with the other observations using a third order spline interpolation.

Since dithering was performed on source, sky subtraction could be 
accomplished with the subtraction of AB pairs of observations.
To obtain optimal sky subtraction, ``super skys" were created by taking the median of the 2 or 3 A or B observations
closest in time to each B and A observation, respectively.  The sky subtracted data cubes were aligned spatially on the
peak flux of the [OIII] line map of the galaxy and co-added.   


\subsection{Kinematic maps}

To measure kinematics from the final co-added data cube, Gaussian functions were fit to 
the [OIII] line in the spectrum of each spatial pixel in which the galaxy was detected.  The resulting rotation velocity ($V$), 
velocity dispersion corrected for instrumental broadening ($\sigma$), [OIII] line, and signal-to-noise 
maps are shown in Figure~\ref{fig:SWIFT}.   The maps have been smoothed spatially with a Gaussian
$\sigma$ of 1 pixel (1 spaxel = 0.235\arcsec), which is much less than the image FWHM of $\sim 1$\arcsec, to 
decrease pixel-to-pixel noise.  A signal-to-noise threshold of 2.5 was applied to these maps.

Also shown in Figure~\ref{fig:SWIFT} are ground-based images from the Canada-France-Hawaii
Telescope (CFHT)/CFH12k camera at $B$, $R$, and $I$, and a $V$-band image from $Hubble$/Advanced
Camera for Surveys (ACS).  The most direct comparison for the SWIFT observations is to the CFHT images, 
rather than the $Hubble$ image, since the image FWHM for the CFHT images is $\sim0.7$\arcsec, similar to that
during the SWIFT observations ($\sim 1$\arcsec).
Figure~\ref{fig:SWIFT} shows that the morphology of the Eagle galaxy does not vary
significantly from observed $B$ to $I$ in the CFHT images.  The SWIFT O[III] linemap has a similar morphology to the CFHT
broad-band images, but is less irregular-shaped, consistent with a poorer image FWHM.
The number of spatial elements per resolution element (i.e., the oversampling rate) for the
SWIFT maps is $\sim 4$.

\begin{figure}
\includegraphics[scale=1.9]{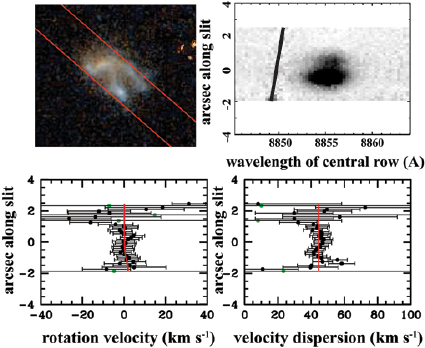}
\caption{$Hubble$ and DEIMOS observations of the Eagle galaxy (top) and fits to the kinematics (bottom).
In the upper left panel is a colour $Hubble$ image created from $V$ and $I$-band
exposures with the position of the 1\arcsec\ wide DEIMOS slit marked.  It is oriented such that north
is up and east is to the left.  The upper right panel shows the [OIII] line in the slit spectrum.    
Th spectrum is shown as observed, so constant wavelength runs diagonally (black line).
The lower panels show the rotation and dispersion profiles created by Gaussian fits to each row of pixels in the [OIII] line.
The black points are data used, and the green points are rejected.  
Note that not all points are independent due to the $\sim 0.7$\arcsec seeing.
The solid red lines are the best fit kinematic models described in \S 4.1.  
\label{fig:DEIMOS}}
\end{figure}

The $V$ map is consistent with a maximum value of $V \times$sin($i$) of $25\pm5$ km s$^{-1}$,
where $i$ is the inclination of the galaxy.   This is the maximum rotation spread along a line with a 
position angle of 145\degr (measured east of north).
From a visual inspection of the $Hubble$ images, it is clear 
that the Eagle has isophotes which are not ellipsoidal and an asymmetric three-armed structure.
In addition, the kinematic major axis is not aligned with the morphological major axis. (The direction of
the morphological major axis is the same as the position angle of the slit for the DEEP2 observations; 
see \S 4.5 and Figure~\ref{fig:DEIMOS}.)
Both of these findings are inconsistent with the Eagle being an edge-on disc.  It is more likely that this galaxy is viewed
nearly face-on, i.e. with an inclination $\la 30$\degr.  We refrain from measuring the inclination due to the 
large uncertainties inherent in measuring low inclinations, especially for systems with 
isophotes which are not ellipsoidal.  For inclinations of 30\degr, 20\degr, 10\degr, and 5\degr,
the inclination-corrected $V$ varies systematically from 50--287 km s$^{-1}$, namely
$V = 50, 73, 144,$ and 287 km s$^{-1}$, respectively. 

The $\sigma$ field shows a gradient such that the southern and northern sections of the galaxy have 
typical $\sigma$'s of $35-50$ and $65-75$ km s$^{-1}$, respectively, with uncertainties of 10 km s$^{-1}$.
As mentioned in \S 1, undisturbed disc galaxies in the local Universe have $\sigma$ values that range from
10--35 km s$^{-1}$ and an average value of $\sim$20--25 km s$^{-1}$ \citep{ghasp} which
are due to the relative motions of individual gas clouds in spiral arms or a thick disc.  The 
much higher $\sigma$ values found for the Eagle likely represent effective velocity dispersions caused by the blurring of 
velocity gradients on scales at or below the seeing limit \citep{wei1, kas7}.  These may not even have
a preferred plane.  In addition, the northern edge of the Eagle, where the highest $\sigma$ values occur, is where
the three ``arms" of the galaxy which are visible in the $Hubble$ images 
meet, possibly indicative of tidal disturbances there.

For a rotating disc galaxy observed at high redshift, seeing (i.e., beam smearing) smooths
out the $V$ gradient and produces a strong peak in the centre of the $\sigma$ map \citep[e.g.,][]{wei1,covi}.  
The $\sigma$ peak is produced in the central
parts of the galaxy where the velocity gradient is strongest and 
seeing smears the light from gas at different velocities together.  There is no seeing-induced peak observed in the $\sigma$ field
of the Eagle.  This is likely due to a combination of its low value of $V \times$sin($i$) and correspondingly shallow 
velocity gradient in its inner parts, and low spatial resolution.  For an image FWHM of $\sim 1$\arcsec,
the resulting decriment in $V$, which is typically $\sim 10-15$ km s$^{-1}$ for a galaxy the size of the 
Eagle \citep{wei1, kas7}, is significantly less than the range in $V$ allowed due to the uncertainty in 
the inclination ($\pm \sim 95$ km s$^{-1}$).  Due to this, the low spatial resolution of the 3D maps, 
and the misalignment of the kinematic and morphological major axes,
we refrain from creating a detailed kinematic model for this system.

\subsection{Comparison with kinematics of galaxies at the same epoch}

\begin{figure}
\includegraphics[scale=0.8]{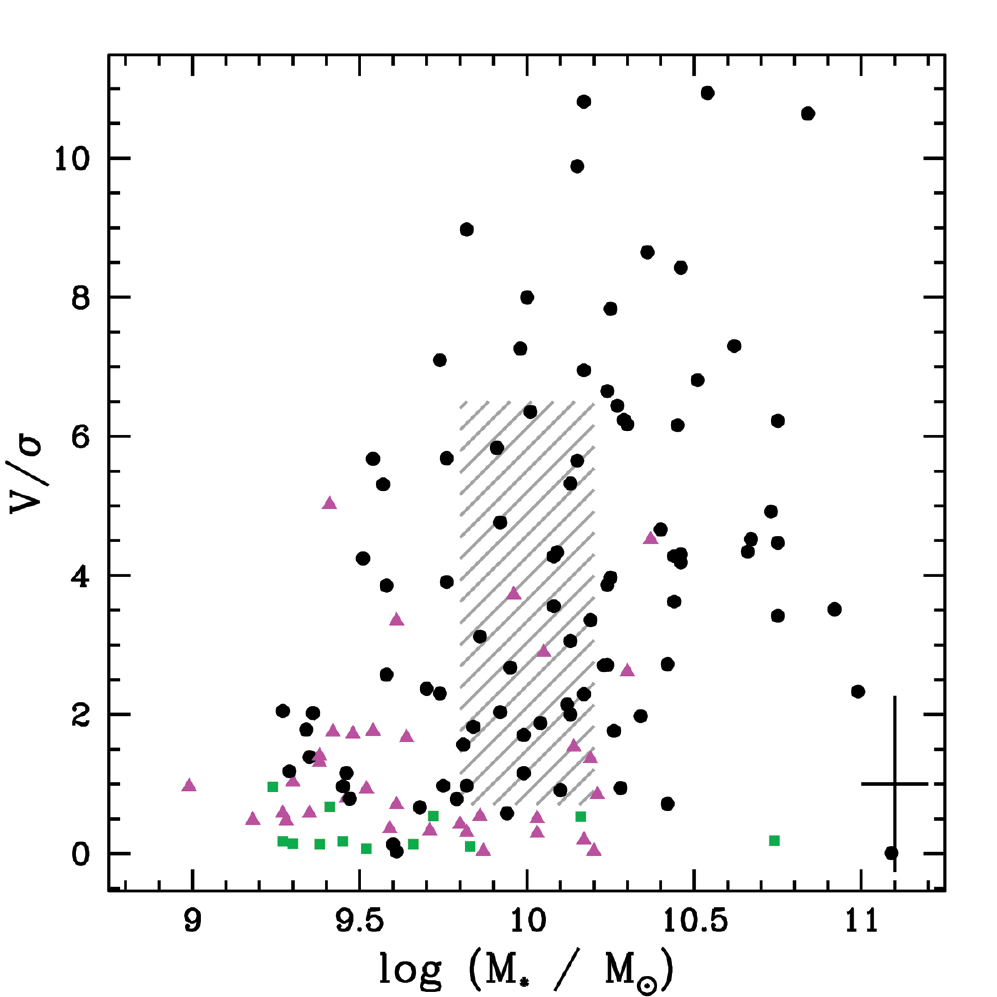}
\caption{The range of values allowed for the ratio of rotation velocity to integrated velocity dispersion 
of the Eagle (grey shading) is compared to values for emission line galaxies at $0.65 < z < 0.85$ from \citet{kas7}
as a function of stellar mass.  The range in $V/\sigma$  for the Eagle is mainly due
to the uncertainty in its inclination.  Other galaxies are coded according to their visual morphology
as deduced from $Hubble$ images by \citet{kas7}: normal (black points), disturbed
(magenta triangles), and compact (green squares).  A typical error bar for these galaxies is shown in black.
Although the Eagle has a large uncertainty in $V/\sigma$, it is not among the very
$V$-dominated systems at this epoch due to its large average $\sigma$ which, at its lowest
possible value of 45 km s$^{-1}$, is still greater than that found for local galaxies.
\label{fig:velsig}}
\end{figure}

In Figure~\ref{fig:velsig}, the kinematics of the Eagle are compared with those of a sample of
galaxies at the same epoch in terms of the ratio of $V$ to $\sigma$, as measured in \citet{kas7}.
These galaxies are essentially selected on emission line strength and are $\ga 80$\% complete down to 
$\sim 10^{9.5}$ M$_{\odot}$ \citep{kas7}.
For the Eagle, the range of allowed values of $V$ from \S 3.1, and an
average $\sigma$ over the entire galaxy of $55 \pm 10$ km s$^{-1}$, are adopted.
Galaxies in Figure~\ref{fig:velsig} are coded according to the visual morphological classification from \citet{kas7}.
It is apparent that galaxies with low $V/\sigma$ values are generally classified as
disturbed, whereas those with high values have normal disc-like morphologies.
Galaxies in the IMAGES Survey at $z\sim0.6$ show a similar range in $V/\sigma$ \citep{puec7}.
Although the Eagle has a large allowable range in $V/\sigma$ due to the large
uncertainty in its inclination, it cannot be among the very $V$-dominated
systems because of its large $\sigma$ which, at its lowest
possible value of 45 km s$^{-1}$, is still greater than that found for local galaxies (\S 1). 
This is consistent with the Eagle's morphological classification as disturbed or major merger (\S 2.2).  

\begin{table*}
\caption{Properties of the Eagle galaxy.}
\begin{tabular}{llllllllllllll}
\hline
$V$ sin($i$) & $\sigma$ & SFR (total) & $M_*$ &$M_B$ & $U-B$ & $12 + log (O/H)$ & $R_{I,Petrosian}$\\
km s$^{-1}$ & km s$^{-1}$ & M$_{\odot}$ yr$^{-1}$ & log M$_{\odot}$ &AB &AB & &$\arcsec$\\
\hline 
$25 \pm 5$ & $35-75 \pm 10$ & $26.3 \pm 0.4$ & $10.0 \pm 0.2$ & $-21.51 \pm 0.08$ & $0.46 \pm 0.09$ & $8.66 \pm 0.05$ & $1.2 \pm 0.2$\\ 
\hline
\end{tabular}
\end{table*}

\section{Multi-wavelength data and spectral energy distribution from AEGIS}

A few IFS studies have incorporated multi-wavelength data which span from the UV or optical 
to the far-infrared \citep[e.g.,][]{fuen,hamm,puec9,puec10}.  
Such data facilitate the formulation of more complete pictures of the galaxies by
allowing for more accurate determinations of e.g., star-formation
rates, dust content, and sometimes metallicity.

Multi-wavelength photometry and a flux calibrated optical spectrum come from the AEGIS Survey.
The spectrum and redshift are from the DEEP2 Survey, a portion of which is incorporated into AEGIS.
A far-infrared flux (observed $24 \mu$m) is adopted from the FIDEL Survey  (Dickinson et al. in preparation) 
which overlaps AEGIS and has a greater depth than the AEGIS $24 \mu$m imaging.
Data are taken from the following telescope/instrument combinations: Keck-2/DEep Imaging 
Multi-Object Spectrograph (DEIMOS) for an optical spectrum, {\it Galaxy Evolution Explorer} for ultraviolet photometry, 
{\it Hubble Space Telescope}/Advanced Camera for Surveys (ACS) for
optical imaging, Canada-France-Hawaii Telescope (CFHT)/CFH12K for optical imaging and photometry,
Palomar/Wide Field Infrared Camera (WIRC) for $K_s$-band photometry, $Spitzer$/Infrared Array Camera (IRAC) for mid-infrared photometry, 
$Spitzer$/Multiband Imaging Photometry for Spitzer (MIPS) for 24\micron\, photometry,
and Chandra/AXAF CCD Imaging Spectrometer (ACIS) 
for counts in the hard and soft bands.  Data reduction is discussed in \citet{aegis} for AEGIS and \citet{samir} for FIDEL.

Photometry was performed on these images as follows, and all measure total galaxy flux.
For the CFHT images, $R$-band magnitudes were measured in circular apertures of radius 
$3 r_g$, where $r_g$ is the $\sigma$ of a Gaussian fit to the image profile.  To derive $B$ and $I$-band magnitudes,
$B-R$ and $R-I$ colours were measured in a 1\arcsec\ radius aperture \citep{coil}.  
The resulting $B$ and $I$-band magnitudes differ from magnitudes measured within $3r_g$ only if there are 
significant colour gradients in the CFHT images, which there are not (Figure~\ref{fig:SWIFT}).
For the WIRC images, photometry was performed using a Kron-like
aperture \citep{bund}.  For the IRAC images, photometry was performed in a 3\arcsec\
diameter aperture \citep{aegis}.  For MIPS and GALEX images, since the point spread functions (PSFs)
are 6\arcsec\ and 5\arcsec\, respectively, PSF fluxes were extracted.

\begin{figure}
\includegraphics[scale=1.3]{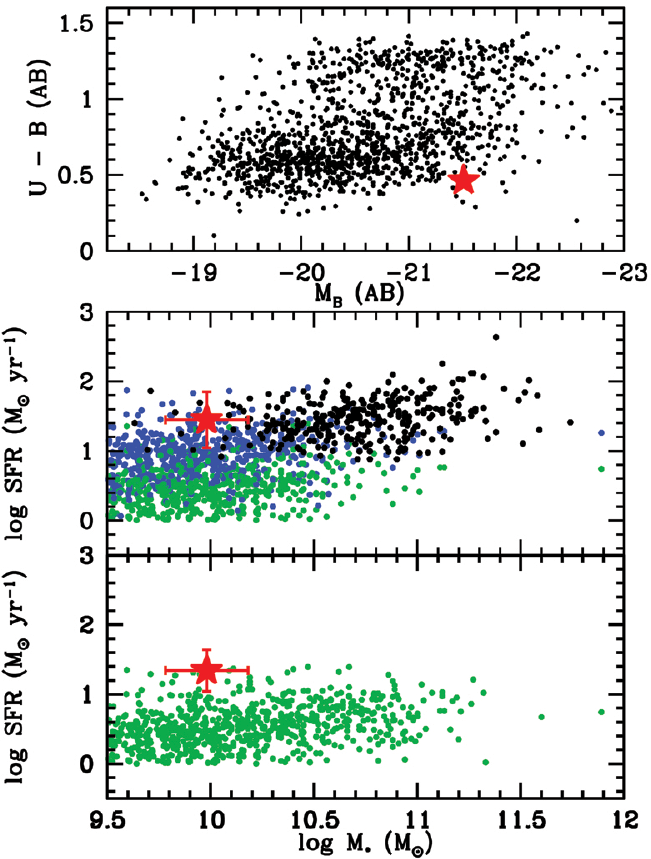}
\caption{The Eagle galaxy (red star) is compared with galaxies 
at $0.65<z<0.85$ in the AEGIS survey (points) in colour-magnitude and
SFR-$M_*$ space.  Top: A colour-magnitude diagram (not corrected for dust) shows that the Eagle 
is on the very bright and blue end of the distribution of galaxies at similar redshifts.
Uncertainties on magnitudes are 10\%.
Middle: A SFR-$M_*$ diagram where the SFR is derived from line emission
for blue galaxies with F$_{24um} < 60$ $\mu$Jy with and without correcting 
for extinction (blue and green points, respectively), and
from line emission and far-infrared flux for galaxies with F$_{24um} > 60$ $\mu$Jy
(black points).  Uncertainties on SFRs and $M_*$'s are 0.2 dex.
The Eagle has a high total SFR for its $M_*$.
Bottom:  This is the same as the middle plot except only the
uncorrected SFRs derived from emission lines are plotted.  The Eagle
has a high SFR as derived from emission lines.
\label{fig:SFR}}
\end{figure}

K-corrections are applied following \citet{cnaw} to obtain rest-frame $U$ and $B$-band magnitudes,
which are given in Table 1.  A rest-frame colour-magnitude diagram in Figure~\ref{fig:SFR}
shows all galaxies in AEGIS at $0.65 < z < 0.85$.  It shows that the Eagle galaxy is extremely bright 
and blue compared to its contemporary galaxies.  

\subsection{Single slit kinematics}

We look to kinematics derived from the DEEP2 slit spectrum for complementarity and completeness.
The [OIII] emission line in this spectrum is shown in Figure~\ref{fig:DEIMOS}.  The image FWHM is
$\sim 0.7$\arcsec.  The resolution was $R\sim5000$ and the slit width was 1\arcsec, which resulted in a spectral resolution of 
FWHM=$0.56$ \AA, or 22 km $s^{-1}$ at the redshift of the Eagle (Gaussian $\sigma=8$ km s$^{-1}$). 
The slit was oriented at the position angle of the photometric major axis which is $49\degr$ 
(east of north; upper left panel of Figure~\ref{fig:DEIMOS}; measured in \S4.5), 
which is approximately perpendicular to the gradient of the velocity field in Figure~\ref{fig:SWIFT}.

As in the IFS data, the spatially extended [OIII] emission line in the galaxy spectrum is used to measure gas rotation and 
dispersion profiles, but at 2D instead of 3D, and under the assumption of a spatially constant $\sigma$.  
The {\tt ROTCURVE} fitting procedure of \citet{wei1} is used to fit a kinematic model.  Briefly, this procedure 
fits Gaussians to each row of pixels in the emission line 
to obtain profiles of $V$ and $\sigma$ along the slit, and rejects discrepant values with automatic
criteria (Figure~\ref{fig:DEIMOS}, bottom panels).  It then measures the light distribution along the slit
and fits a Gaussian to it.  Finally, {\tt ROTCURVE} fits models of the position-velocity
distribution along the slit, taking the seeing into account.  The model has two parameters: $V$ (not
corrected for inclination) and $\sigma$.
The resulting $V$ and $\sigma$ for the Eagle are $0 \pm 9$ km s$^{-1}$ and $45 \pm 6$ km s$^{-1}$, respectively.

The 2D slit spectrum shows no evidence of rotation, whereas the 3D velocity field
shows a rotation gradient of $25 \pm 5$ km s$^{-1}$.  This is because the slit was placed approximately 
perpendicular to the kinematic major axis.  
The $\sigma$ measured from the slit spectrum is shown to be constant along the slit, as
demonstrated by the Gaussian fits to each row of the emission line (black points in the
bottom left panel of Figure~\ref{fig:DEIMOS}).  The slit spectrum does not show the $\sigma$ gradient found 
in the 3D $\sigma$ map because it is luminosity-weighted and not as deep as the 3D observations.
Therefore, it is biased towards the brightest regions of the galaxy (as depicted in the [OIII] line map in 
Figure~\ref{fig:SWIFT}), where the $\sigma$ is 55$\pm 10$ km s$^{-1}$.

This comparison between 2D and 3D spectroscopy of the same object demonstrates the 
power of 3D spectroscopy to reveal the full kinematics of high redshift galaxies.

\subsection{Star-formation rate}

Historically, there are two main approaches to determine the star-formation rate (SFR) of a galaxy.  
These depend on the wavelength of the data at hand: optical or 
UV which trace the output of hot young stars (but which need to be corrected 
for interstellar extinction), and far-infrared which acts as a calorimeter \citep[e.g.,][]{kenn}.
As multi-wavelength observations become more commonplace, studies of galaxy
SFRs increasingly combine UV/optical and infrared indicators to obtain better 
accuracy \citep[e.g.,][]{dale, calz, ken09}.
These studies take advantage of the fact that dust mostly radiates in the 
infrared and sub-mm, while intercepting radiation at UV and optical wavelengths.  Therefore, 
the addition of a UV or optically-determined SFR (not corrected for dust) and an infrared-determined SFR gives a more
accurate account of the total SFR of a system.  Furthermore, a comparison of the two SFRs
gives insight into dust content.  

Measurements of SFRs from far-infrared data rely on stellar
models to predict the energy output from stars, and assume that most of the radiation from young stars
is extincted and re-processed into the infrared.
A SFR is derived from the $24\mu$m flux of the Eagle following \citet{kai} using spectral energy distribution (SED) 
templates from \citet{char}.  A \citet{krou} initial mass function is assumed, which does not differ
significantly for the purposes of this paper from a \citet{chab} IMF \citep{riek}. The resulting SFR for 
$F_{24\mu \rm m} = 71.8 \pm 2.6$ $\mu$Jy is 6 M$_{\odot}$  yr$^{-1}$, which has
a factor of $\sim 2 $ uncertainty.  This is consistent with SFRs derived with the formulations of \citet{riek}
and \citet{daleh}.  


SFRs derived from nebular emission lines (like H$\beta$ which is used here) rely on stellar models to predict
the ionizing luminosity of O stars.  They assume that the extinction of the recombination photons is given by
the Balmer decriment (i.e., H$\beta$/H$\alpha$).  A calibration from \citet{kenn} for
H$\alpha$ line luminosity is typically adopted: SFR (M$_{\odot}$ yr$^{-1}$) = $7.9 \times 10^{-42}$ $L$(H$\alpha$) (ergs s$^{-1}$). 
The resulting SFR is for a \citet{salp} IMF and is converted to a \citet{chab} IMF by a multiplicative
factor of 0.66 \citep{riek}.  



An H$\beta$ line luminosity is adopted from the DEEP2 spectrum since the Eagle is 
compared with galaxies in DEEP2 for which measurements and calibrations are performed in the same manner.
The H$\beta$ line luminosity is measured following \citet{ben07}, and is $10^{41.9 \pm 0.1}$ ergs s$^{-1}$.

A {\it lower limit} to the emission line-determined SFR of the Eagle is derived from the H$\beta$ line
under the assumption of {\it no extinction} (i.e., case B recombination at $10,000$\degr K: H$\beta$/H$\alpha$ = 0.35, \citealt{oster}),
namely 11.3 M$_{\odot}$ yr$^{-1}$. An {\it upper limit} is derived 
using a lower bound to the Balmer decriment for a distribution of blue galaxies in DEEP2 over $0.33<z<0.39$
(the redshift range where both lines can be measured) from \citet{ben07}, namely 0.11.  This results in a 
SFR of 36.0 M$_{\odot}$ yr$^{-1}$.  Finally,  to estimate the {\it mean} SFR of the Eagle from its 
H$\beta$ line luminosity, the average Balmer decriment for galaxies in \citet{ben07} is adopted,
namely 0.198.  This results in a SFR of 20.3 M$_{\odot}$ yr$^{-1}$.  In summary, the SFR of the Eagle
derived from H$\beta$ and not corrected further for extinction is 20.3 M$_{\odot}$ yr$^{-1}$ for a \citet{chab} IMF, with lower and upper limits of 
11.3 and 36.0 M$_{\odot}$ yr$^{-1}$, respectively.  In this paper, we adopt the mean SFR since we know
there is some dust present due to its detection at $24\mu$m.

SFRs derived from UV flux rely on stellar models to predict the UV flux from O and B stars and assume
that the extinction is given by a measurement of the UV slope.  Because intermediate-age stars also contribute to the UV flux,
a simultaneous fit is typically made to the young and older stellar populations, and a dust model is assumed.
Such a fit for the Eagle is adopted from \citet{samir} and results in a SFR of $9.4 \pm 2.0$ M$_{\odot}$ yr$^{-1}$ (corrected for extinction with the UV slope).  The uncorrected value is 3.5 M$_{\odot}$ yr$^{-1}$ .  The discrepancy between this measurement and the H$\beta$-derived SFR is somewhat
unusual.  Along with the low 24$\mu m$-derived SFR, it is evidence that extinction is low in star-forming regions in the Eagle.

The SFR from the 24$\mu$m flux (6 M$_{\odot}$  yr$^{-1}$) is less than
even the lower limit to the SFR derived from H$\beta$ which is not corrected for extinction (11.3 M$_{\odot}$ yr$^{-1}$).
Even though both these measurements are uncertain to within a factor of $\sim 2$ due to
statistical scatter in the flux-to-SFR conversion, they differ systematically.  The SFR derived
from 24$\mu$m is calculated under the assumption that dust captures all the UV flux (calorimeter assumption),
and the SFR derived from H$\beta$ with zero dust correction (11.3 M$_{\odot}$  yr$^{-1}$, lower limit) assumed
that dust captures none of the ionizing or H$\beta$ flux.  For the vast majority of galaxies in the DEEP2 Survey,
the SFR derived from 24$\mu$m is greater than the uncorrected line-derived SFR by a factor of $\sim 3$.
There are hardly any galaxies where the uncorrected emission line-derived SFR is greater than the 24$\mu$m-derived SFR.
The Eagle is an exception as its emission-line SFR is greater by at least a factor of $\sim 2$ (11.3 versus 6 M$_{\odot}$  yr$^{-1}$).
Similarly, \citet{bell05} find the 24$\mu$m-derived SFR dominates the uncorrected UV-derived SFR.

This discrepancy between 24$\mu$m and uncorrected H$\beta$-derived SFRs
suggests that the fraction of line emission escaping the sites of star-formation is large so
that the calorimeter assumption is violated and the extinction of H$\alpha$ could be lower than normal.
In addition, a low dust content is consistent with the blue colour of the galaxy, uncorrected for dust (Figure~\ref{fig:SFR}, top panel).
A total SFR for the Eagle is estimated by adding the emission line (not corrected for dust) and
24$\mu m$-derived SFRs, and results in 26.3 M$_{\odot}$  yr$^{-1}$, 
which has a factor of 2--3 uncertainty.   

To compare the Eagle with galaxies at similar redshifts,  
in Figure~\ref{fig:SFR} the SFRs of galaxies in the AEGIS Survey at $0.65 < z < 0.85$ are
plotted versus stellar mass (following \citealt{kai}), and the Eagle galaxy is highlighted.  
The sample is $> 80$\% complete for $M_*\ga 10^{9.6}$ M$_{\odot}$
and $>95$\% complete for M$_* \ga 10^{10}$ \citep{kai}. 
These galaxies are from the same data set and have SFRs measured in a homogeneous manner.
Figure~\ref{fig:SFR} shows that the 
Eagle has a high SFR compared to galaxies at similar stellar masses, even when
only the SFR derived from line emission is taken into account.  

\begin{figure}
\includegraphics[scale=1.2]{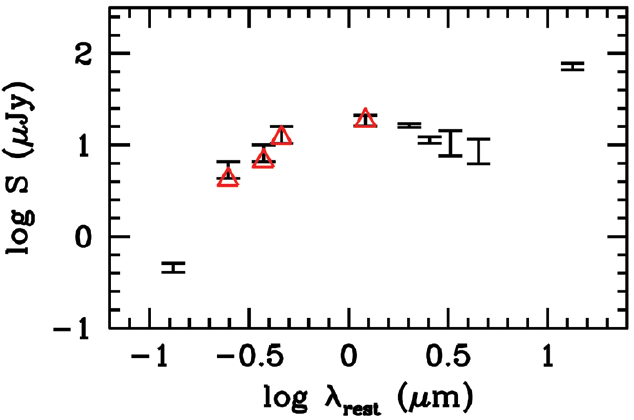}
\caption{Multi-wavelength broad band spectral energy distribution of the Eagle galaxy (black error bars)
and the best-fit galaxy stellar population model to the optical and near-infrared fluxes from \citealt{bund} (red triangles).
The data, from left to right, are $NUV$, $B$, $R$, $I$, $K$, IRAC channels 1, 2, 3, 4, 
and MIPS $24\mu m$ fluxes.   The fluxes are log $S = -0.34\pm 0.05, 0.726\pm0.09, 0.908\pm0.09, 1.111\pm0.09,
1.213\pm0.02, 1.053\pm0.03, 1.018\pm0.14, 0.930\pm0.14,$ and $1.856\pm0.04\, \mu$Jy, respectively.
\label{fig:SED}}
\end{figure}

\subsection{Stellar mass \& broad-band spectral energy distribution}

The multi-wavelength SED for the Eagle galaxy is shown in Figure~\ref{fig:SED}.
Fits of galaxy stellar population models to the observed $B, R, I,$ and $K_s$-band fluxes
were performed by \citet{bund} to obtain a stellar mass estimate ($M_*$) for the Eagle.
{These are the wavelengths at which stellar population models are the best-characterised.}
The galaxy models constituted a grid of synthetic spectral energy distributions from \citet{bc03} which
spanned a range of exponential star-formation histories, ages (restricted to be less than the age of the Universe
at the redshift of the Eagle), metallicities, and
dust contents.  A \citet{chab} initial mass function, which does not differ significantly 
from the \citet{krou} function assumed for the SFR calculation, was adopted.
At each point on the grid of models, the following is calculated:
the $K_s$-band stellar mass-to-light ratio, minimum $\chi^2$, and 
probability that each model accurately describes the galaxy. The corresponding $M_*$ is 
then determined by scaling the stellar mass-to-light ratio to the $K_s$-band luminosity based on \
the total $K_s$-band magnitude. The probabilities are then summed (marginalised) across the 
grid and binned by model stellar mass, yielding a stellar mass probability distribution for
the galaxy.  The median of the distribution is adopted as the best estimate.
The $M_*$ measured in this way is robust to degeneracies in the models, such as 
those between age and metallicity \citep{bund}.
The best-fit model has a log $M_*$ of $10.0 \pm 0.2$ M$_{\odot}$, where the uncertainty
is taken from the width of the probability distribution.  There are additional 
systematic uncertainties associated with the IMF and stellar population synthesis
models adopted which can be as large as 0.4 dex \citep{barr}.  At the redshift of the
Eagle, taking into account the TP-AGB phase of stellar evolution does not produce
significant changes in stellar mass estimates \citep[e.g.,][]{barr}.

We refrain from including the IRAC channel 1 waveband in the SED fit since doing so 
decreases the goodness of the model fit.  This is possibly because of uncertainties
in stellar population models at this wavelength.  The mid to far-infrared
points are not included in the fit because stellar population models do not 
currently include the dust modelling necessary to fit them.
Furthermore, although the best-fit model also provides estimates of age, metallicity, 
star-formation history, and dust content, these quantities are much more affected
by degeneracies and are poorly constrained compared with the stellar mass \citep{bund}.



\subsection{Test for AGN contamination}

In this section, we test for active galactic nuclei, or AGN,
contribution to the emission lines in the spectrum of the Eagle.  This could
affect derived quantities such as kinematics, SFR, and stellar mass.
There are two methods which are regarded as the most effective for the detection of AGN
in high redshift galaxies, where it is generally not possible to isolate the nuclear regions. 
These are optical emission line and X-ray selections.  Neither selection is perfect.  
Emission lines from the AGN can be overpowered by star-formation, and X-rays can be
affected by heavy absorption of gas in close proximity to the AGN.  However, a combination of
these two methods is the best currently available to identify AGN \citep{renb}.

To test for a possible contribution from AGN to 
the Eagle, we first look to an emission line diagnostic diagram
derived for galaxies in the DEEP2 Survey over $0.3 < z < 0.8$ \citep[][Figure 5]{renb}.
This diagram uses integrated rest-frame $U-B$ colour and [OIII]/H$\beta$ line ratio to
separate star-forming galaxies from AGN.    
The Eagle galaxy lies in the star-forming region of the diagram given its [OIII]/H$\beta$ value 
of $0.53\pm0.2$ from the DEEP2 spectrum.

Next, we look to the deep $200$ks Chandra data.  No counts were detected at the 
coordinates of the Eagle in either of the Chandra bands.  This places $5\sigma$ upper limits on the X-ray
luminosity of AGN in the Eagle of L$_X$ = $1.1 \times 10^{-16}$ and $8.2 \times 10^{-16}$ ergs s$^{-1}$ cm$^{-2}$ for the soft and hard bands, respectively.

Furthermore, we note that the near-infrared through IRAC portion of the SED in Figure~\ref{fig:SED}
does not show the typical power law shape of an AGN.  In conclusion, the kinematics and emission 
line luminosities of the Eagle are likely not strongly affected by AGN.

\subsection{Metallicity}

The oxygen abundances of the Eagle and galaxies in AEGIS 
at similar redshifts ($0.65 < z < 0.85$) are estimated.  This is
done by adopting the relation between $R_{23}$
and the gas phase oxygen abundance by \citet{mcg91}, following \citet{kk04}.  
The upper branch of this calibration is adopted because the majority of galaxies in \citet{kk04}
and \citet{kz99} with NII/H$\alpha$ measurements fall there.
This measurement relies on equivalent widths of emission lines.
It works in part because star-forming galaxies have a relatively narrow range of 
continuum shapes between [OII] and [OIII], and do not have large 4000\AA\ breaks.
For the Eagle, rest-frame equivalent widths of the [OII] $\lambda 3727$,
[OIII] $\lambda 5007$, and H$\beta$ lines ($67.00 \pm 1.25$, $77.33 \pm 1.82$, and $28.1 \pm 0.73$, 
respectively  and all in \AA),  result in an oxygen abundance of 12 + log (O/H)  $ = 8.66 \pm 0.05$.

\begin{figure}
\includegraphics[scale=1.2]{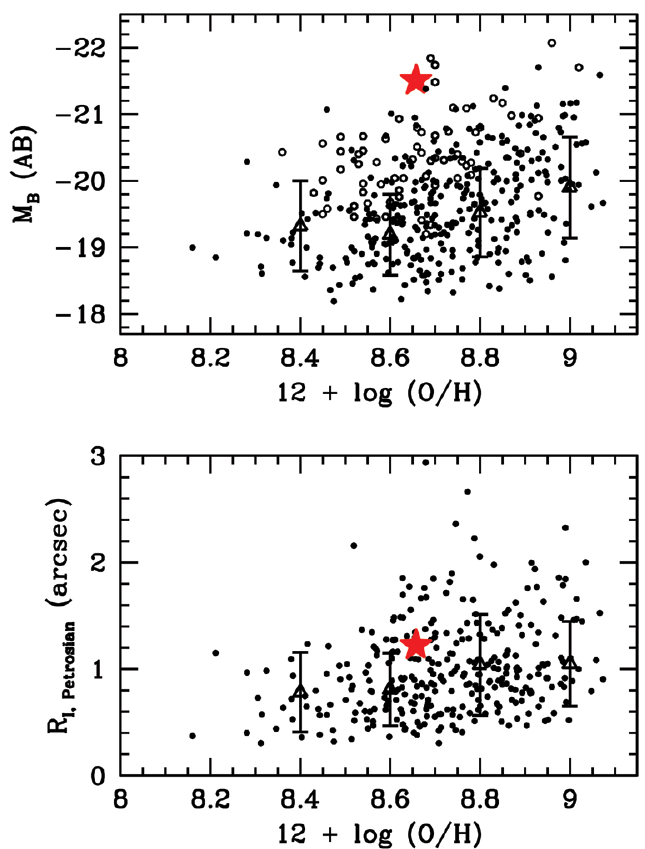}
\caption{Oxygen abundances are compared with rest-frame $B$-band 
magnitudes (top) and $I$-band Petrosian radii (bottom) 
for galaxies in the AEGIS Survey at $0.65<z<0.85$ (filled circles)
and from \citet{kk04} (open circles).  Typical uncertainties in the measurements of oxygen abundances,
sizes, and $B$-band magnitudes are 0.05 dex, 0.1 dex, and 10\%, respectively.
In addition, for the oxygen abundances, there can be systematic errors in the calibration
which may be $> 0.1$ dex \citep{kk04}.
Binned averages for the AEGIS galaxies
are shown as open triangles and error bars show the rms scatter.  The Eagle galaxy is plotted as a red star.
It is bright and large for its oxygen abundance.
\label{fig:Z}}
\end{figure}

In Figure~\ref{fig:Z}, the oxygen abundances of AEGIS galaxies and the Eagle are plotted versus
rest-frame $B$-band magnitudes and Petrosian radii (measured from
$I$-band $Hubble$ images as described below).  
This galaxy sample is complete down to at least $M_B$ (AB) $\la 19$ for blue
galaxies (no red galaxies are shown in the figure) in the redshift range plotted \citep{cnaw}.
Galaxies from \citet{kk04} in the same redshift
range are also compared.  
The Eagle galaxy is brighter by $\sim 2$ $B$-band 
magnitudes and larger by $\sim 0.4$ dex than average for its oxygen abundance.
The oxygen abundance of the Eagle is similar to those of galaxies 
at similar stellar masses in the IMAGES Survey at $z\sim0.6$ \citep{myri}.

\section{A disturbed and highly star-forming galaxy}

The SWIFT 3D kinematic fields demonstrate that the Eagle galaxy has a gradient in $\sigma$
which ranges from $35 - 75 \pm 10$ km s$^{-1}$ and a $V$ of at least $25 \pm 5$ km s$^{-1}$.  
Whatever the process (or processes) is that creates the abnormally high $\sigma$ values, it is likely active throughout most of
the galaxy.  These $\sigma$ values are likely caused by the blurring of velocity gradients, which may
not even have a preferred plane, on scales at or below the seeing limit  \citep{wei1,kas7}.  The high $\sigma$'s are 
consistent with the morphology of the Eagle, which is quantitatively classified as a major merger,
and shows structures unlike those of a normal disc galaxy.

Table 1 summarises the properties of the Eagle galaxy, and Figures 3--6 compare it with galaxies at the same epoch.
Compared to these galaxies, the Eagle is extremely bright and blue in the
rest-frame optical.  It is highly star-forming with an unusually high SFR as derived 
from line emission, and a comparably low SFR as derived from far-infrared luminosity.
This suggests that most of the star-formation in the Eagle occurs in dust-free regions. 
The Eagle is also large  and bright for its metallicity.  These properties are not surprising given that the Eagle was 
primarily selected to have extended and bright emission lines.

The Eagle could be in an early stage of a star-burst brought on by a major merger.  
It may have not yet had enough time
for metal enrichment, or to have formed the mass of dust typically found 
in highly star-forming galaxies.  It is possible that the high SFR could drive winds that expose star-forming 
regions, and, given the lower metallicity, may explain the low extinction.  (There
are local dwarf galaxies which show similar phenomena, but this galaxy is much more massive
than such systems.)  The low metallicity and strong star-burst may also be explained
by a major merger event that brings lower metallicity gas from the outer parts
to the centre of the system \citep[e.g.,][]{kew06}.


It is likely that the mechanism which causes the star burst is the same,
or initiated by, that which causes the high velocity dispersion and disturbed morphology.
This mechanism is likely a major merger.

\section*{Acknowledgments}
The Oxford SWIFT integral field spectrograph is directly supported by a Marie Curie Excellence Grant from the
European Commission (MEXT-CT-2003-002792, Team Leader: N. Thatte).  It is also supported by additional
funds from the University of Oxford Physics Department and the John Fell OUP Research Fund.  Additional
funds to host and support SWIFT were provided by Caltech Optical Observatories.

This paper is based in part on observations obtained at the Hale Telescope at Palomar Observatory
as part of a collaborative agreement between the California Institute of Technology, its
divisions Caltech Optical Observatories and the Jet Propulsion Laboratory (operated for NASA),
and Cornell University.  

Additional support was provided for by the Observational Astrophysics Rolling Grant at Oxford and 
the Oxford Astrophysics PATT Linked Grant ST/G004331/1.  The following NSF grants to the DEEP2 Survey 
are acknowledged: AST00-71198, AST05-07483, and AST08-08133.  

L. Fogarty would like to acknowledge the generous support of the Foley-B\'ejar Scholarship through 
Balliol College, Oxford and the support of the STFC.

This work of LAM was carried out in part at Jet Propulsion Laboratory, California Institute of Technology, 
under a contract with NASA. LAM acknowledges support from the NASA ATFP program.

\end{document}